# Libron–phonon coupling and hydrogen-bond dynamics in the vacancy-ordered perovskite $(NH_4)_2SnCl_6$: a temperature- and pressure-dependent Raman study

Vasco S. Neto[1], Mayra A. P. Gómez[1,2], Bruno S. Araújo[1], Alejandro P. Ayala[1]*

[1] Departamento de Física, Universidade Federal do Ceará, Campus do Pici, Fortaleza, CE 60440-900, Brasil

[2] Departamento de Física dos Materiais e Mecânica (FMT), Instituto de Física da Universidade de São Paulo (IF-USP), São Paulo, São Paulo, Brasil. CEP 05314-970.

## Abstract

Vacancy-ordered perovskites of the $R_2MX_6$ family have drawn renewed interest for their optoelectronic and thermoelectric potential, yet how their A-site cation dynamics couples to the lattice phonons under external perturbation remains largely unexplored. Here we synthesized $(NH_4)_2SnCl_6$ and probed its structure and lattice dynamics as a function of temperature (10–300 K) and pressure (0–10.1 GPa) by single-crystal X-ray diffraction and Raman spectroscopy. The average cubic structure persists over the whole temperature range, with smoothly varying lattice parameters and no structural anomaly; below ≈ 100 K, however, the $[SnCl_6]^{2-}$ modes acquire a libron–phonon renormalization ($E_{\text{eff}} \approx$ 4.7 meV) and a symmetry-selective line asymmetry, while the N–H stretch passes through a non-monotonic minimum near 110–130 K ($E_{\text{eff}} \approx$ 9.3 meV) — both tracking the classical-to-quantum crossover of the ammonium rotor established by neutron scattering and NQR. The ammonium bending and stretching linewidths, by contrast, are governed by pure dephasing against the low-energy librational and translational manifold rather than by anharmonic decay, and carry no crossover signature. Under pressure the internal $[SnCl_6]^{2-}$ phonons stiffen sublinearly with no discontinuity, whereas the cavity subsystem responds twice: the symmetry-allowed external translational $F_{2g}$ mode of $NH_4^+$ gains Raman intensity above ≈ 1.3 GPa, and the N–H stretch inverts its pressure slope near 1.7 GPa as the hydrogen bonds pass from shortening to bending. No change of space group accompanies these features, and the ambient spectrum fully recovers on decompression. Temperature and pressure together show that the rigid $[SnCl_6]^{2-}$ framework is essentially decoupled from a dynamically active $NH_4^+$ subsystem, whose libration and hydrogen bonding carry the response to both perturbations — a decoupling that offers a route to tune cation–phonon coupling independently of the octahedral network.

* Corresponding author: ayala@fisica.ufc.br

## I. INTRODUCTION

Vacancy-ordered perovskites are usually derived from the traditional $A_2B_2X_6$ double perovskite. In these materials half of the B-site ions are missing, leading to an octahedral lattice classified as a 0D perovskite type [1,2]. 0D perovskites are characterized by isolated inorganic octahedra surrounded by isolated cations that act as spacers between the octahedral units [3]. This 0D structure enables the manifestation of intrinsic properties of individual metal halide units, such as highly efficient photoluminescence and broad emission bands with long decay times, and promotes the formation of self-trapped excitons (STEs) — excitons that are rapidly trapped, possibly dragging a local lattice distortion with them [1,4,5,6,7]. Because the octahedra are isolated rather than corner-sharing, the A-site cation in these structures is comparatively free to reorient within its cage, and this cation dynamics — librational, rotational, or, for protonated cations, tunneling motion — couples directly to the phonons of the surrounding lattice. In the wider hybrid-perovskite family, this cation–phonon coupling is recognized as a central factor shaping carrier mobility and charge-carrier scattering [8]. Vacancy-ordered perovskites, in which the cation and octahedral subsystems are structurally decoupled, offer a comparatively clean platform for isolating this coupling from the octahedral-tilting effects that dominate in 3D perovskites.

Despite this potential, the coupling between cation dynamics and lattice phonons in vacancy-ordered perovskites remains far less explored than in the corner-sharing 3D hybrid perovskites, especially regarding its response to external perturbations such as pressure. For the ammonium-templated members of the $A_2SnX_6$ family, particularly $(NH_4)_2SnCl_6$, previous studies have focused primarily on molecular rotations and on possible couplings between lattice phonons and librational states over a wide temperature range [9]. The highly stable and symmetric structure of $(NH_4)_2SnCl_6$ shows no structural phase transition between room temperature and 10 K, as confirmed by the measurements reported here, making it an ideal system for studying the rotational dynamics of the ammonium ion within a rigid crystalline lattice. Techniques such as inelastic neutron scattering [10], nuclear magnetic resonance [11], and Raman spectroscopy [12] have been employed to investigate these motions, revealing a transition from a classical to a quantum regime, marked by the splitting of rotor states below 100 K [13]. Punkkinen *et al.* [14] proposed a model describing the coupling between lattice vibrations and tunneling states [15], but this model is only applicable below 50 K, whereas the effects reported in earlier studies were observed around 100 K — an unresolved gap between the available theory and the observed temperature window. Pressure offers an independent, and for this family largely untested, route to perturb the same coupling: a recent high-pressure study of the isostructural vacancy-ordered perovskite $Rb_2TeBr_6$ found that compression alone drives a continuous octahedral reorientation with direct consequences for the optical response, well before any crystallographic phase transition is reached [16], indicating that the pressure dimension is an open and active question for the broader vacancy-ordered family. Whether an analogous

response — a reorientation of the lattice, a perturbation of the $[NH_4]^+$ tunneling states, or both — occurs in $(NH_4)_2SnCl_6$ under pressure has, to our knowledge, not been addressed.

This renewed interest in vacancy-ordered perovskites is part of a broader resurgence of attention toward halide perovskites as next-generation energy materials, owing to their outstanding optoelectronic properties — tunable bandgap, high absorption coefficient and carrier mobility, long diffusion length, long carrier lifetime, and defect tolerance [17,18,19]. This combination of properties enables a wide range of technological applications. Within this broader family, the $A_2SnX_6$ subclass crystallizes in the vacancy-ordered double-perovskite structure and combines the advantage of nontoxicity relative to other metals with good stability under atmospheric exposure [20]. $(NH_4)_2SnCl_6$ belongs to this family and has been extensively studied in recent years for different technological purposes. Given its good tolerance for defects and ion replacement, several doping studies have been conducted on this material: Te-doped compounds exhibit highly efficient orange emission centered at 590 nm [21], while Sb-doped samples show near-infrared emission at 734 nm with a large Stokes shift under 360 nm excitation, and a yellow emission at 590 nm under 390 nm excitation [22]. The replacement of the ammonium ion, or use of its hydrated forms, has also been extensively investigated for optoelectronic applications [2].

Since temperature and pressure are fundamental thermodynamic variables that modulate lattice dynamics in complementary ways, addressing this gap requires a combined view of how $(NH_4)_2SnCl_6$ behaves across both conditions. Phonon dynamics, including the cation–phonon coupling discussed above, directly influence properties such as specific heat and charge transport, and in turn the band structure and optoelectronic response of the material. More specifically, both the inorganic and the organic sublattice vibrations govern the optical response: the inorganic modes drive carrier localization and trapping, while the organic ones broaden the emission spectrum [23]. In this work, we investigate the temperature dependence of the cell parameters and of the vibrational spectra of $(NH_4)_2SnCl_6$ by single-crystal X-ray diffraction and Raman spectroscopy, and extend this analysis to high-pressure conditions to address the open question identified above. A comprehensive assignment of the vibrational modes is also presented, providing a deeper understanding of the mechanisms governing vibrational dynamics in this material.

## II. METHODS AND SYNTHESIS

### A. Synthesis

Single crystals of $(NH_4)_2SnCl_6$ were grown by the slow-evaporation method. Commercial reagents were used without further purification: hydrochloric acid (HCl, 47 wt% in $H_2O$), ammonium chloride ($NH_4Cl$), and tin(II) chloride ($SnCl_2$), purchased from Sigma-Aldrich and Alfa Aesar. Stoichiometric amounts of $NH_4Cl$ and $SnCl_2$ were dissolved in HCl in a 1:1 molar ratio. The mixture was stirred at 60 °C until completely dissolved, then left undisturbed in a beaker covered with a perforated plastic film to allow slow solvent evaporation, during which Sn(II) is oxidised to Sn(IV). Single crystals suitable for X-ray diffraction formed after 24 hours.

**B. Single-crystal X-ray diffraction**

Single-crystal X-ray diffraction data were collected using φ and ω scans on a Bruker D8 Venture diffractometer equipped with a PHOTON II CPAD detector and an Incoatec IµS 3.0 Mo Kα microfocus source (λ = 0.71073 Å). Temperature-dependent experiments were conducted between 90 and 300 K using an Oxford Cryostream 800 Plus attached to the diffractometer. Diffraction patterns were collected every 3 K in the 90–150 K range and every 10 K over the remainder of the range.

Unit-cell determination and data collection were performed with the APEX4 software suite (Bruker AXS Inc., 2021). Data reduction and global cell refinement used the Bruker SAINT+ package, and multi-scan absorption correction was applied via SADABS [24]. Using OLEX2 [25] as the graphical interface, the structure was solved by the intrinsic phasing method implemented in ShelXT [26], which located most non-hydrogen atoms. The remaining atoms were located from difference-Fourier maps calculated over successive full-matrix least-squares refinement cycles on $F^2$ using ShelXL [27], and refined with anisotropic displacement parameters. Structure validation and geometrical analysis were carried out with PLATON [28]. Crystal-structure artwork was prepared with VESTA [29].

**C. Raman spectroscopy**

Raman spectra were recorded using a Jobin-Yvon T64000 triple spectrometer in backscattering geometry, coupled to an Olympus BX41 microscope with a 20× long-working-distance achromatic objective. The 514.5 nm line of an $Ar^+$ laser was used as the excitation source. For the temperature-dependent measurements, the sample was cooled with a closed-cycle helium cryostat, with temperature controlled to within 0.1 K using a Lakeshore 330 controller. Spectral parameters were obtained by fitting Voigt profiles, either with the Fityk software [30] or, for the temperature-dependent mode positions, with an in-house iterative fitting routine (“backfit”) in which a background and a librational contribution are refined alternately. The background is the anharmonic (Balkanski) baseline for the internal $[SnCl_6]^{2-}$ modes and the measured lattice contraction $a$(T) for the N–H stretch; the librational contribution is a two-level population term $S \cdot \tanh(E_eff/2k_BT)$. The effective librational energy $E_{eff}$ is taken from the model fit to the full data set, and its uncertainty from a case bootstrap: 1000 replicates, resampling the mode positions with replacement and, for the N–H stretch, the $a$(T) data as well, so that the uncertainty of the contraction background is propagated. The full estimator — multistart over several seeds followed by a joint refinement — is applied to each replicate. We quote the full width at half maximum of the resulting distribution as the uncertainty range; the reasons for preferring it to the formal fit error are given in Sec. III C.

For the high-pressure measurements, pressure was generated using a diamond-anvil cell (DAC) fitted with stainless-steel gaskets pre-indented to a thickness suitable for a sample chamber of approximately 150 μm in diameter, sealed by tightening the cell's screws. The chamber was loaded with powdered sample together with Nujol as the pressure-transmitting medium. Pressure inside the chamber was determined from the luminescence lines of $Cr^{3+}$ in small ruby chips placed alongside the sample, using the IPPS-Ruby2020 pressure scale [31].

## III. RESULTS AND DISCUSSION

### A. Crystal structure

Single-crystal X-ray diffraction analysis at room temperature confirms that $(NH_4)_2SnCl_6$ crystallizes in a cubic lattice, space group $Fm\bar{3}m$, with lattice parameter $a$ = 10.0519(4) Å, consistent with the structure reported by Brill *et al.* [32]. The structure can be described in terms of two interpenetrating sublattices (Figure 1). In the anionic sublattice, Sn atoms occupy the highly symmetric $4a$ Wyckoff position ($O_h$ point symmetry) at the corners and face centers of the unit cell, with each Sn atom octahedrally coordinated by six Cl atoms at the $24e$ position ($C_{4v}$ point symmetry). Since the octahedra share no vertices, the electronic structure of the inorganic framework is essentially that of a confined molecular ion, which is the basis for classifying these materials as "0D" perovskites. The cationic sublattice is composed of ammonium ions occupying the $8c$ Wyckoff position, with tetrahedral ($T_d$) site symmetry; these ions sit in the cuboctahedral cavities formed by the surrounding chloride ions. Unlike the simple spherical cations (e.g., $Cs^+$) found in analogous compounds such as $Cs_2SnCl_6$, the ammonium ion has internal structure and orientational degrees of freedom.

The interaction between the ammonium cation and the surrounding chloride cage is governed by electrostatic interactions and hydrogen bonding. The four N–H bonds of the ammonium ion are oriented toward the faces of the surrounding octahedra (Figure 1a) rather than directly at individual chloride ions; each hydrogen atom is equidistant from three chloride ions, forming a trifurcated (three-centered) hydrogen bond (N–H···3Cl). This is fundamentally different from the linear hydrogen bonds found in simpler salts such as $NH_4Cl$, where the proton sits in a deep, narrow potential well directed at a single acceptor. In the trifurcated bond the potential surface is flatter and more extensive, allowing greater librational freedom.

To investigate potential phase transitions or lattice distortions, temperature-dependent single-crystal XRD experiments were conducted between 90 and 300 K. No discontinuities in the unit-cell parameters were observed over the entire range, indicating that the cubic symmetry is preserved throughout (Figure 1b,c).

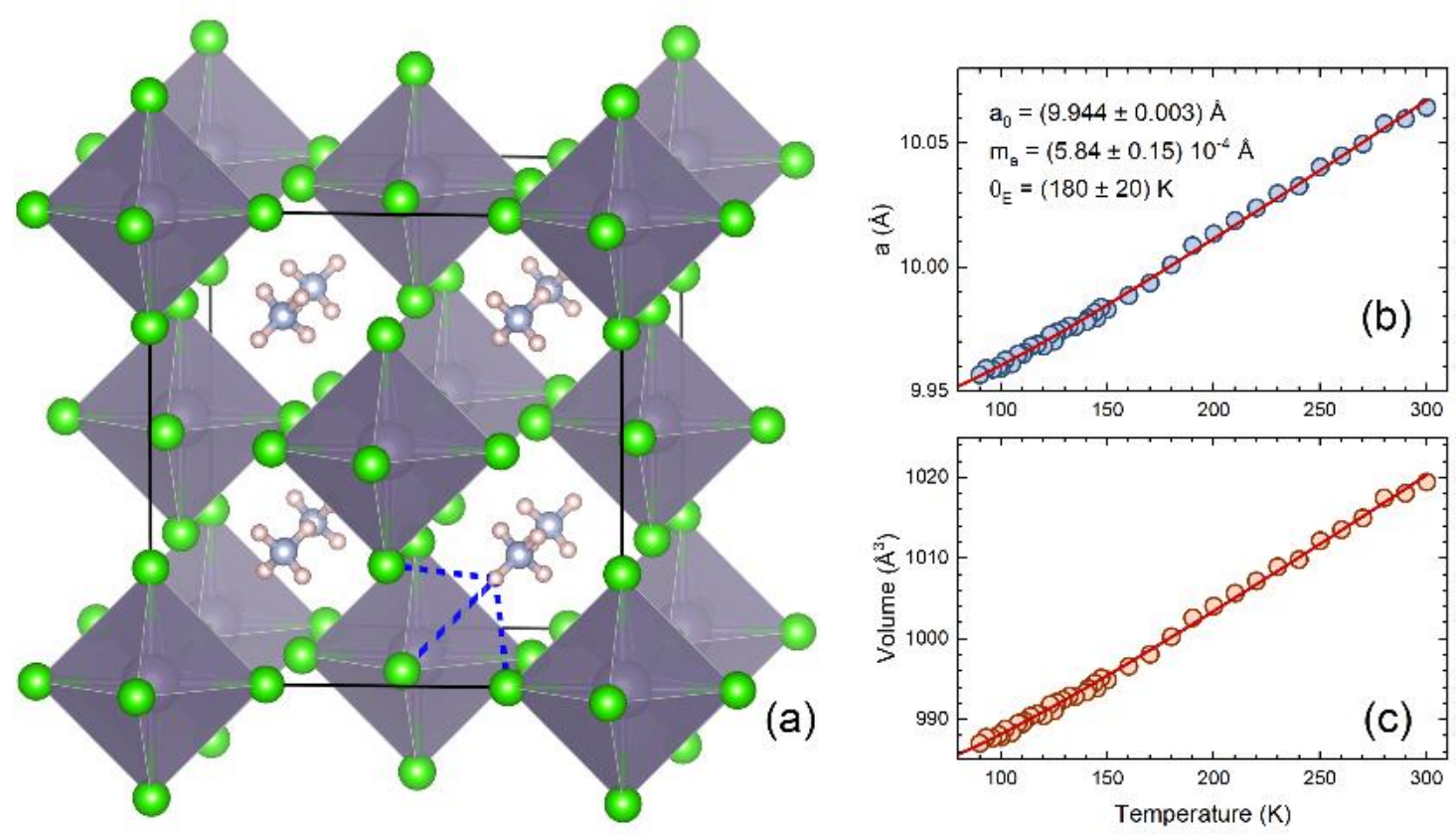

**Figure 1.** (a) Cubic unit cell of $(NH_4)_2SnCl_6$ ($Fm\bar{3}m$), with isolated $[SnCl_6]^{2-}$ octahedra and $NH_4^+$ ions in the cuboctahedral cavities; for one ammonium ion, blue lines mark the trifurcated N–H···3Cl hydrogen-bond contacts of a single H atom to its three nearest $Cl^-$ neighbours (H···Cl = 2.93 Å, N–H···Cl = 134.6°, from the 100 K structure). (b) Temperature dependence of the lattice parameter $a$ and (c) of the unit-cell volume $V$ between 90 and 300 K; solid lines are fits to the Einstein–Grüneisen expression (Eq. 4), with fit parameters as annotated in (b).

## B. Raman-active modes and group-theory assignment

Raman spectroscopy is a sensitive probe of subtle interatomic interactions and modifications in local symmetry in crystalline structures. For the cubic lattice of $(NH_4)_2SnCl_6$, group-theory analysis predicts 18 Raman-active vibrational degrees of freedom, distributed among the irreducible representations of the $O_h$ factor group at the Γ point of the Brillouin zone [33] as $2A_{1g} \oplus 2E_g \oplus 4F_{2g}$. Because the molecular sub-units share no atoms, the factor group can be decomposed into individual contributions: $[SnCl_6]$: $A_{1g} \oplus E_g \oplus F_{2g}$, and $[NH_4]$: $A_{1g} \oplus E_g \oplus 3F_{2g}$. The Raman spectrum of $(NH_4)_2SnCl_6$ thus separates into three distinct regions: (i) a low-wavenumber region (100–400 cm$^{-1}$), containing lattice vibrations and internal $[SnCl_6]^{2-}$ motions; (ii) a mid-wavenumber region (1300–1700 cm$^{-1}$), corresponding to ammonium bending modes; and (iii) a high-wavenumber region (2800–3300 cm$^{-1}$), dominated by ammonium stretching modes.

Within the low-wavenumber region, three characteristic modes of the $[SnCl_6]^{2-}$ octahedron are observed: the non-degenerate symmetric stretch $\nu(A_{1g})$, the doubly degenerate asymmetric stretch ν(Eg), and the triply degenerate asymmetric bend $\delta(F_{2g})$. As expected, the wavenumbers follow the order $\delta(F_{2g}) < \nu(E_g) < \nu(A_{1g})$ [34], located at 174, 239 and 319 cm$^{-1}$, respectively. The observed wavenumbers closely match those reported for the isostructural, same-anion $K_2SnCl_6$ (172, 244 and 324 cm$^{-1}$ for $\delta(F_{2g})$, $\nu(E_g)$ and $\nu(A_{1g})$; [35]), confirming that the ammonium cation perturbs the octahedral force field only weakly.

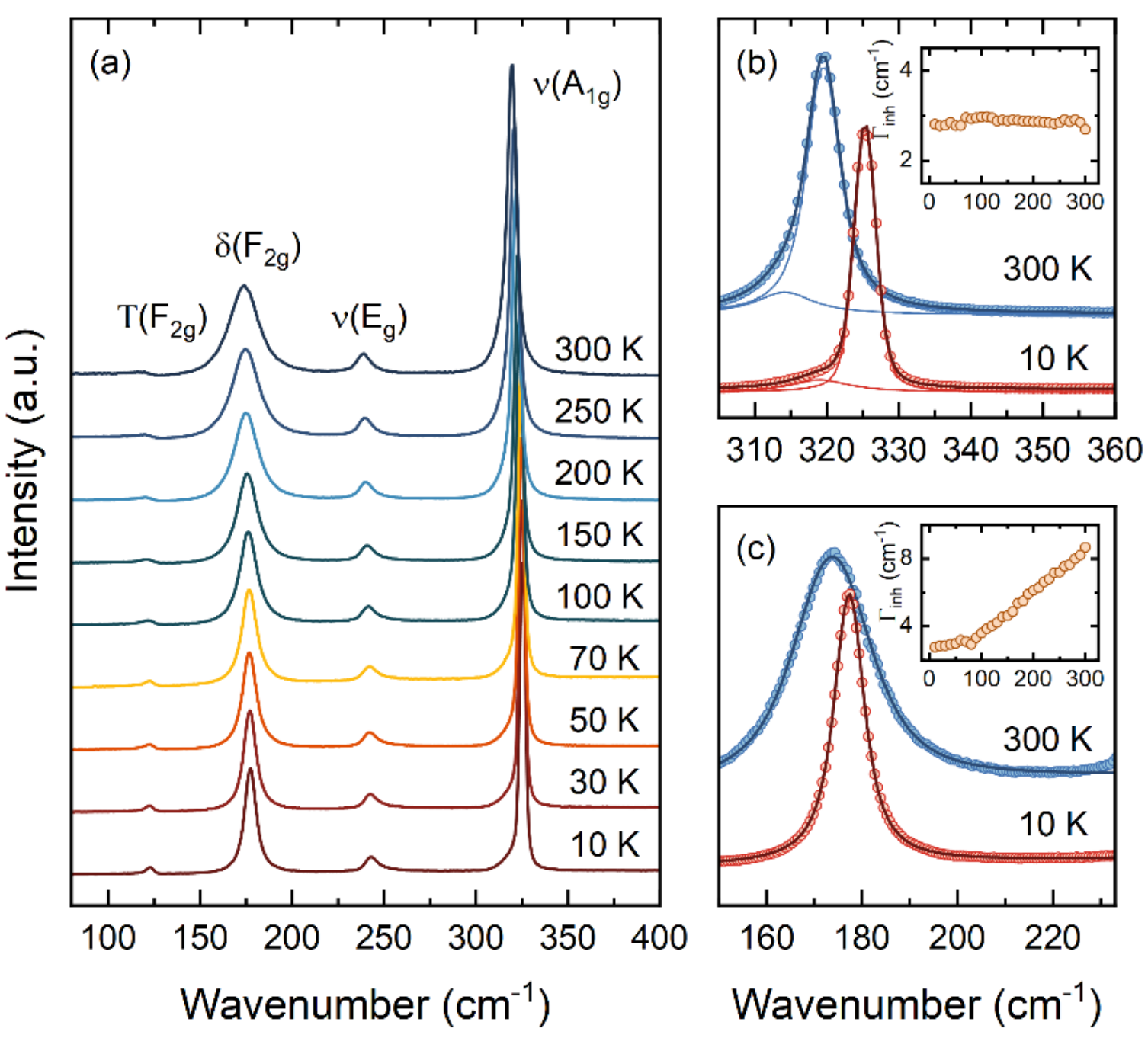

**Figure 2.** Temperature-dependent Raman spectra and line-shape decomposition of the low-wavenumber modes of $(NH_4)_2SnCl_6$. (a) Stacked spectra from 10 to 300 K, showing the four Raman-active low-wavenumber modes: the external translational $F_{2g}$ mode of $NH_4^+$ [$T(F_{2g})$, ≈120 cm⁻¹], the octahedral bend $\delta(F_{2g})$, the stretch $\nu(E_g)$, and the symmetric stretch $\nu(A_{1g})$. (b) $\nu(A_{1g})$ mode at 300 K (red) and 10 K (green) fitted with a Voigt plus a Lorentzian, the additional Lorentzian accounting for the asymmetric second component on the low-wavenumber side; inset, temperature dependence of the Gaussian (inhomogeneous) width $\Gamma_{inh}$ of the Voigt component, which is small (≈2.8 $cm^{-1}$) and nearly temperature-independent. (c) $\delta(F_{2g})$ mode at 300 K (red) and 10 K (green) fitted with a single Voigt; inset, $\Gamma_{inh}(T)$ of that Voigt, which increases monotonically from ≈3 to ≈9 $cm^{-1}$ on warming.

The decomposition above also predicts one additional Raman-active mode for the ammonium ion beyond its four internal vibrations ($\nu_1$–$\nu_4$, which account for the $A_{1g}$, Eg, and two of the three $F_{2g}$ components). In the free-ion $T_d$ point group, translations of the ion transform as $F_2$ and rotations (librations) as $F_1$; correlating these up to the crystal's $O_h$ factor group, $F_2$ maps onto the Raman-active $F_{2g}$ or IR-active $F_{1u}$ representations, whereas $F_1$ maps onto $F_{1g}$ or $F_{2u}$, both of which are silent in $O_h$ (neither has a matching linear or quadratic basis function). The remaining $F_{2g}$ mode is therefore assigned to the rigid-body translation of the $NH_4^+$ ion within its chloride cage — an external, translational lattice mode — rather than to its libration; the assignment of the extra low-wavenumber $F_{2g}$ mode to an A-site cation translation is standard for $R_2MX_6$ antifluorites and related cubic halide perovskites [35,36]. By the same argument, the $NH_4^+$ librations central to the rotational/tunneling dynamics discussed above are expected to be Raman-silent at this site symmetry, which is consistent with why inelastic neutron scattering, NMR, NQR, and dielectric spectroscopy — rather than Raman — have historically been the tools used to probe them [10,11,12,13]. The librational motion of the $[SnCl_6]^{2-}$ octahedron is likewise Raman-inactive: it transforms as $F_{1g}$, which carries no linear or quadratic basis function and is therefore silent, so the octahedral rotary mode seen by neutron scattering (Sec. III C) does not appear directly in the Raman spectrum. The same rigid-ion lattice-dynamics analysis of the $R_2MX_6$ antifluorites confirms this assignment, computing the octahedral libration as a silent $F_{1g}$ rotational mode inaccessible to Raman or infrared spectroscopy [35].

**C. Temperature-dependent Raman spectroscopy**

Figure 2(a) shows the temperature evolution of the low-wavenumber region. As temperature decreases, the three $[SnCl_6]^{2-}$ modes harden with no change in their number or relative positions; in addition, a weak low-wavenumber $F_{2g}$ mode near 120 $cm^{-1}$, barely visible at room temperature, gains intensity on cooling (see below). The modest anharmonicity of these modes is characteristic of the rigid, essentially isolated octahedra of the vacancy-ordered halides [37,38]. Each mode was fitted with a Voigt profile so as to separate the Lorentzian FWHM ($\Gamma_L$) — the lifetime-limited, homogeneous width that enters the anharmonic analysis — from the Gaussian width, which was monitored as a diagnostic of inhomogeneous broadening with the independently measured instrumental resolution held fixed. The temperature evolution of the peak position and $\Gamma_L$ is shown in Figure 3(a)–(d) for $\nu(A_{1g})$ and $\delta(F_{2g})$, and in Figure S1 for $\nu(E_g)$ and the translational $F_{2g}$ mode.

Whereas the $\delta(F_{2g})$ bending mode retains a symmetric line shape over the whole temperature range (Figure 2c), the $\nu(A_{1g})$ stretching mode is asymmetric, with a second component on its low-wavenumber side that is increasingly resolved on cooling; a comparable asymmetry is discernible in $\nu(E_g)$, which is however too weak for a reliable two-component decomposition. For $\nu(A_{1g})$ this was reproduced by fitting the main line with a Voigt profile and the low-wavenumber shoulder with an additional Lorentzian (Figure 2b). The mode selectivity constrains its origin: $\nu(A_{1g})$ is non-degenerate and cannot split through a lifting of degeneracy by a local symmetry lowering, so its asymmetry must arise either from coupling to an additional degree of freedom or from a distribution of slightly different local environments; conversely, the triply degenerate $\delta(F_{2g})$ mode, which could split under a non-cubic local distortion, remains symmetric. This is the opposite of the pattern expected from static degeneracy lifting, and instead points to a symmetry-selective coupling that is strongest for the totally symmetric stretching motion [39]. Because the asymmetry sharpens on cooling — precisely when the lines narrow and band overlap is reduced — it cannot be an artefact of unresolved neighboring bands, and is identified as an intrinsic effect. The modes are further distinguished by their Gaussian (inhomogeneous) width: those of the two stretching modes are nearly temperature-independent ($\approx$ 2.8 $cm^{-1}$ for $\nu(A_{1g})$, Figure 2b inset, and $\approx$ 5.5 $cm^{-1}$ for $\nu(E_g)$), whereas that of the symmetric $\delta(F_{2g})$ mode grows monotonically from ≈ 3 cm⁻¹ at 10 K to ≈ 9 cm⁻¹ at 300 K (Figure 2c, inset) — a distinct channel that broadens the δ($F_{2g}$) line without making it asymmetric.

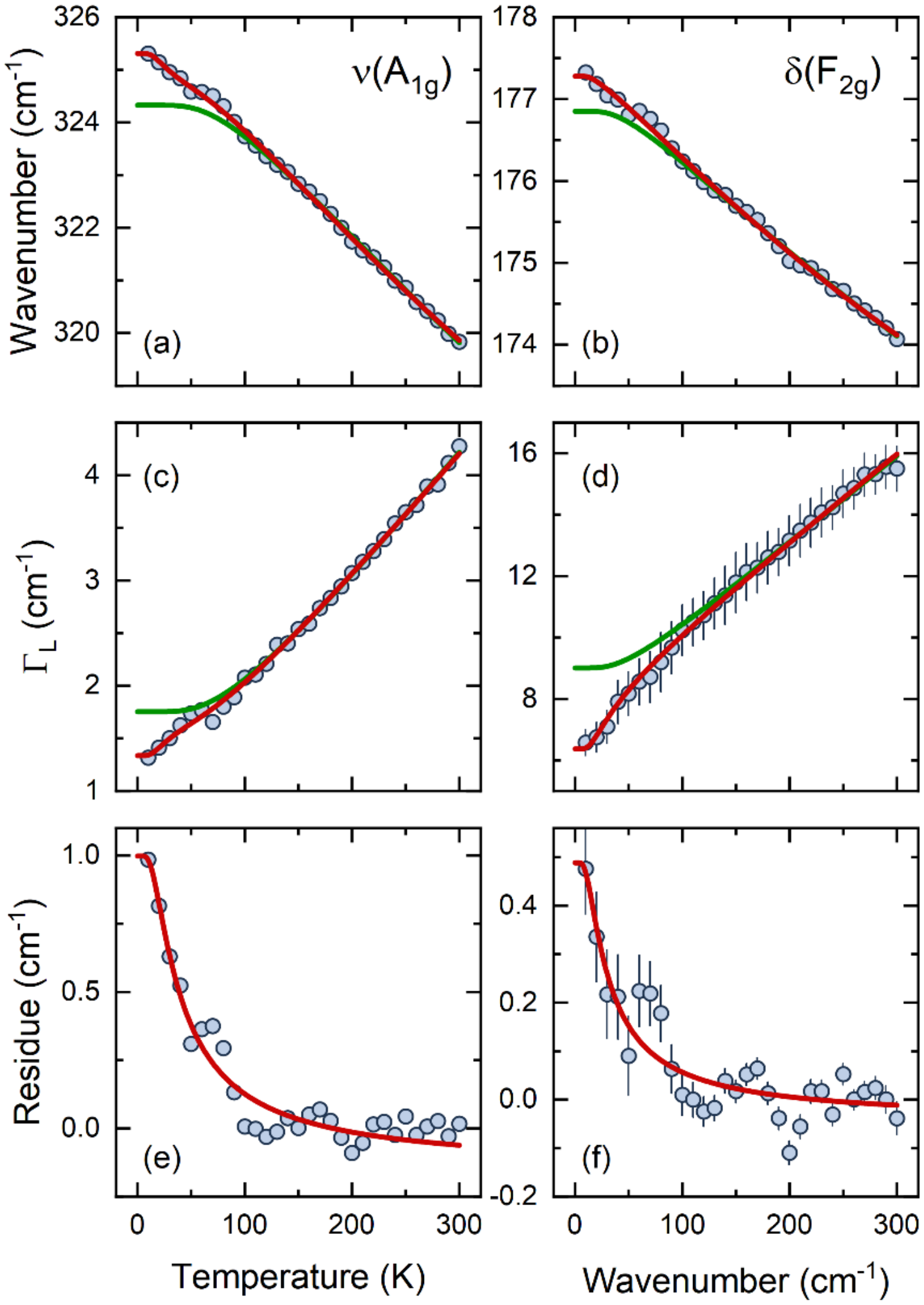

**Figure 3.** Temperature dependence of the wavenumber (a, b), Lorentzian width $\Gamma_L$ (c, d), and residual deviation from the anharmonic baseline (e, f) of the ν($A_{1g}$) (left) and δ($F_{2g}$) (right) modes. Green line, anharmonic (Balkanski) model (Eqs. 1–2); red line, anharmonic-plus-libron model (Eqs. 1–3) in (a)–(d), and the libron term (Eq. 3) fitted to the residual in (e), (f). Both modes deviate systematically from the anharmonic baseline below ≈ 100 K, the deviation being larger for ν($A_{1g}$) than for δ($F_{2g}$). The corresponding data for ν(Eg) and for the external translational $F_{2g}$ mode of $NH_4^+$, omitted here for clarity, are given in Figure S1.

In the absence of phase transitions, the temperature dependence of phonon position and FWHM can be described by the Balkanski model [40], which accounts for anharmonic lattice contributions to the phonon self-energy:

$$\omega(T) = \omega_0 + A\left[1 + \frac{2}{e^x - 1}\right] + B\left[1 + \frac{3}{e^y - 1} + \frac{3}{(e^y - 1)^2}\right] \quad (1)$$

$$\Gamma(T) = \Gamma_0 + C\left[1 + \frac{2}{e^x - 1}\right] + D\left[1 + \frac{3}{e^y - 1} + \frac{3}{(e^y - 1)^2}\right] \quad (2)$$

with $x = \frac{\hbar\omega_0}{2k_BT}$, $y = \frac{\hbar\omega_0}{3k_BT}$, and $A$, $B$, $C$, $D$, $\omega_0$, $\Gamma_0$ as fitting parameters. A temperature-independent residual term $\Gamma_0$ is retained in the linewidth (Eq. 2): without it, the bare Bose factor overestimates the thermal broadening of the $[SnCl_6]^{2-}$ modes by roughly a factor of two and the anharmonic fit fails. The quartic (four-phonon) term $D$ in Eq. (2) was fixed to zero: at low temperature it is degenerate with $\Gamma_0$ and $C$ (all three contribute to $\Gamma(0) = \Gamma_0 + C + D$), and releasing it does not improve the fit over the measured range. Position and width were fitted independently, because the position is determined two to three orders of magnitude more precisely than the width and a joint fit would be dominated by the former.

Figure 3(a)–(d) shows that the position and width of the $\nu(A_{1g})$ and $\delta(F_{2g})$ modes deviate systematically from the anharmonic model below ≈ 100 K, the deviation growing on cooling; the same behavior is found for $\nu(E_g)$ (Figure S1). For the well-resolved $\nu(A_{1g})$ mode the position renormalization reaches ≈ 0.9 $cm^{-1}$ at 10 K; for the $\delta(F_{2g})$ mode it is smaller (≈ 0.5 $cm^{-1}$), consistent with weaker coupling. A purely anharmonic description therefore fails below the onset, leaving the systematic residual shown in Figure 3(e),(f).

The magnitude of the deviation below 100 K is well described by adding to the anharmonic baseline a libron–phonon term whose temperature dependence follows the population difference of an effective two-level librational system — a lower state that couples strongly to the $[SnCl_6]^{2-}$ phonons and an upper one that couples weakly, separated by $E_{\mathrm{eff}}$ — so that the amplitude $S$ measures the contrast between the two coupling strengths,

$$\Delta\omega(T) = S\tanh\left(\frac{E_{\mathrm{eff}}}{2k_BT}\right) \quad (3)$$

with an analogous term for $\Gamma_L(T)$. This contribution saturates at low temperature and vanishes above an onset temperature set by $E_{\mathrm{eff}}$, and is shown as the deviation $\delta\omega(T)$ from the anharmonic baseline in Figure 3(e) for $\nu(A_{1g})$ and Figure 3(f) for $\delta(F_{2g})$. A global fit of the three inorganic modes sharing a common $E_{\mathrm{eff}}$ reproduces the data, and fitting the modes individually returns values consistent with it; independent fits to the mode positions and to the linewidths also yield mutually consistent energies. The libron amplitude follows the symmetry hierarchy $\nu(A_{1g})$, $\nu(E_g) > \delta(F_{2g})$, the bending mode being only weakly affected — consistent with the absence of a satellite component in its line shape (Figure 2). For the well-resolved $\nu(A_{1g})$ mode the model fit yields $E_{\mathrm{eff}} \approx 4.7$ meV, with an onset near 90–100 K and a bootstrap FWHM range of 3.5–6.2 meV (Sec. II C). The ν(Eg) mode returns a consistent value (≈ 5.7 meV, FWHM 3.2–8.2 meV; Figure S1), corroborating $\nu(A_{1g})$; only the $\delta(F_{2g})$ mode is too weakly coupled to constrain $E_{\mathrm{eff}}$ — its bootstrap distribution collapses in about a third of the replicates and its lower bound is unresolved — consistent with the absence of a libron satellite in its otherwise symmetric line shape.

Two features of these fits justify quoting a bootstrap range rather than the formal fit error. First, the reduced $\chi^2$ of the position fits ranges from ≈1 to ≈120 across the modes, so the parameter errors returned by the peak-fitting step are inconsistently scaled and, for the strongest modes, underestimate the true scatter by more than an order of magnitude; case resampling is insensitive to this because it does not use those error bars. Second, because the libron term saturates below ≈50 K, the information on $E_{eff}$ is carried by the curvature of that saturation, so the quadratic approximation behind the covariance estimate does not hold. This sensitivity is intrinsic to a saturating two-level term rather than a limitation of the sampling, and is what the bootstrap quantifies: replicates that under-sample the saturation region returns large $E_{eff}$, producing the right-skewed distributions of Figure S2. Their 95 % percentile intervals are correspondingly wide (3.7–17.8 meV for ν($A_{1g}$)) and would not discriminate between the energy scales compared below, so we quote the FWHM, which is insensitive to that tail and measures the width of the distribution rather than being a confidence interval. Because the $NH_4^+$ libration is itself Raman-silent at this site symmetry (Sec. III B), it is not observed directly; the effect reported here is its indirect signature in the renormalization of the Raman-active $[SnCl_6]^{2-}$ modes to which it couples.

A fourth, weak $F_{2g}$ mode is present at low wavenumber (≈122 cm$^{-1}$ at 10 K, softening to ≈117 cm$^{-1}$ at 290 K), barely visible at room temperature but gaining intensity on cooling. Following the group-theory decomposition (Sec. III B), it is the external translational $F_{2g}$ lattice mode of the $NH_4^+$ ion within its chloride cage — a rigid-body translation of the cation — the same mode that gains Raman intensity under pressure (Sec. III D); this assignment matches the translational lattice mode identified in $R_2MX_6$ antifluorites and related cubic halide perovskites [35,36]. Independent first-principles support comes from a recent density-functional perturbation theory study of $(NH_4)_2SnCl_6$, which places an $F_{2g}$ mode at 110 cm$^{-1}$ whose displacement pattern is dominated by the $NH_4^+$ sublattice, alongside the three internal $[SnCl_6]^{2-}$ modes computed at 176, 228 and 309 cm$^{-1}$ [23] — an ordering that coincides with the assignment adopted here. In the rigid-ion analysis of $K_2SnCl_6$ this cation-translational $F_{2g}$ mode falls at 72 cm$^{-1}$ [35]; its shift to ≈120 cm$^{-1}$ in $(NH_4)_2SnCl_6$ is consistent with the lighter mass of $NH_4^+$ (18 u vs 39 u for $K^+$) and with the additional restoring force of the N–H···Cl hydrogen bonds, both of which raise the wavenumber. Being too weak for a stable Voigt decomposition, it was fitted with a single Lorentzian, so no inhomogeneous (Gaussian) width is available for it (Figure S1). Its position and width are reproduced by the anharmonic Balkanski model with only a marginal libron contribution — a residual of ≈0.3 cm$^{-1}$ below ≈50 K: of the four modes it is the most anharmonic and the least renormalized by the libron, as expected for a rigid-body translation that couples only weakly to the $NH_4^+$ libration. Its intensity relative to ν($A_{1g}$) roughly doubles between room temperature and 10 K, mirroring the pressure-induced intensity gain of the same mode (Sec. III D); both reflect the tightening of the chloride cavity around the ammonium ion — by cooling and by compression, respectively.

The onset temperature (≈90–100 K) coincides with the classical-to-quantum crossover of the $[NH_4]^+$ rotor reported from inelastic neutron scattering (INS) and NQR [10,11], and with the temperature below which the $[NH_4]^+$ stretching mode departs from its high-temperature behavior (Figure 4). Notably, the average cubic structure remains smooth across this range — the lattice parameters follow a single linear trend with no anomaly (Figure 1b,c) — which identifies the crossover as a dynamical rather than a structural effect, consistent with the absence of any phase transition. The effective energy from the Raman renormalization (≈4.7 meV) is about three times smaller than the first librational transition of the $[NH_4]^+$ ion measured by INS (13.4 meV; [10]). We do not read this as a revision of the neutron value, which is independently supported — by a subsequent single-crystal neutron study on the deuterated analogue $(ND_4)_2SnCl_6$ [41], by the dispersionless character of the librational level, and by the consistency of the librational energy with the reorientational barrier (≈590 K) from NQR [10,11]. Rather, the two probes report different quantities: INS measures the bare 0→1 librational transition, whereas the Raman renormalization reflects the effective energy scale that governs the phonon self-energy. Because Eq. (3) collapses the full librational manifold — several excited states, each further split by rotational tunnelling — onto a single effective two-level splitting weighted by how strongly each state couples to the phonon, $E_{\mathrm{eff}}$ is expected to fall below the bare 0→1 transition rather than to coincide with it. A contribution from a lower-energy degree of freedom, such as the soft rotary (librational) mode of the $[SnCl_6]^{2-}$ octahedra reported by INS in this family [41], may add to the same effect. The present Raman experiment thus provides a temperature-resolved, in-situ probe of a libron–phonon coupling previously inferred only from neutron scattering.

The high-wavenumber region is entirely composed of ammonium-related phonons. The first two phonons, at 1404 and 1665 cm$^{-1}$, are bending vibrations, while the other two, at 3157 and 3238 cm$^{-1}$, correspond to N–H stretching vibrations (Figure 4). A weak, poorly resolved band system between ≈ 3260 and ≈ 3390 cm$^{-1}$ is assigned to overtones and combinations of the bending modes and is not analyzed further.

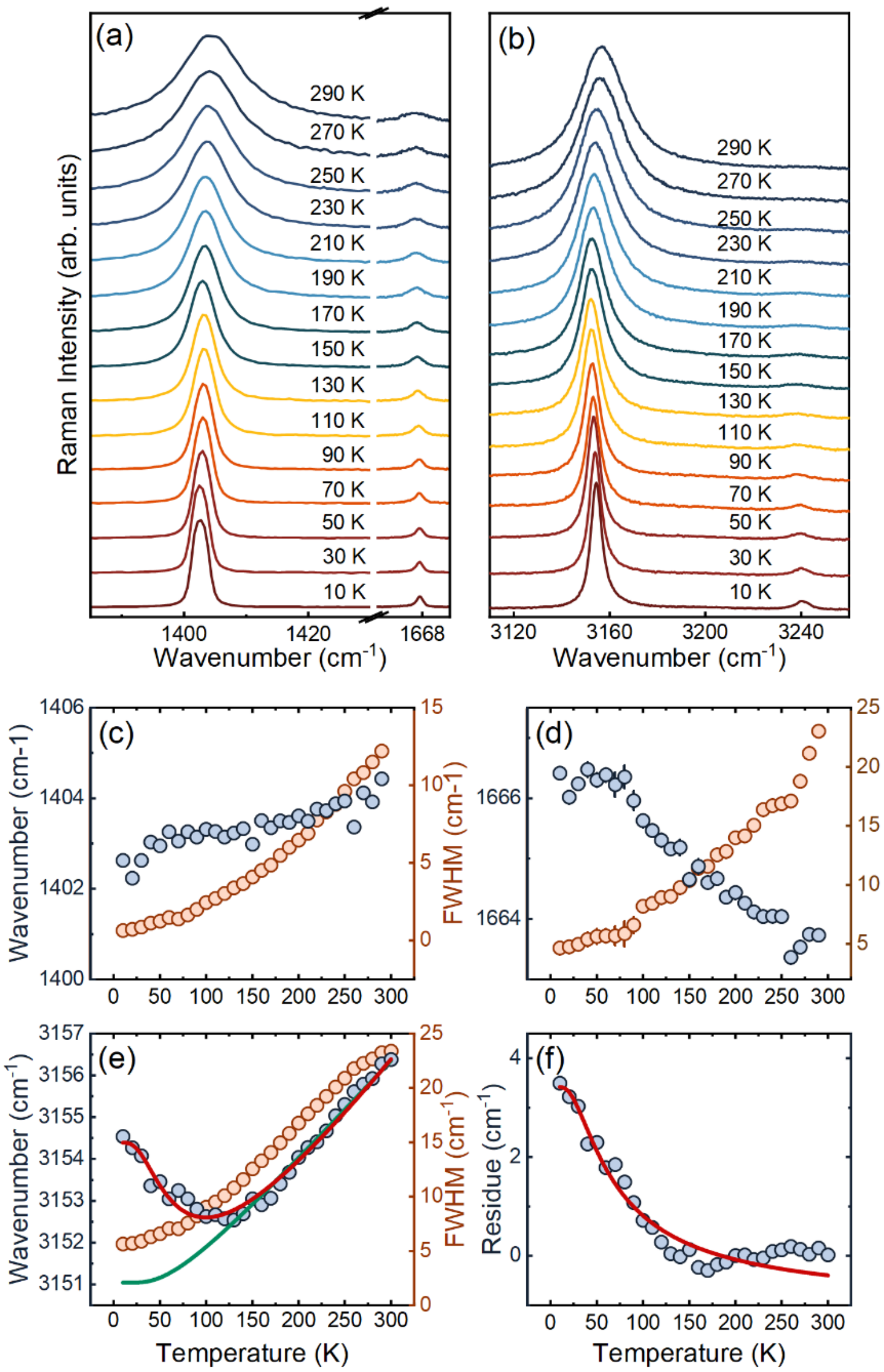

**Figure 4.** Temperature dependence of the ammonium-ion Raman modes. (a) Bending region, showing the $\nu_4(F_2)$ mode near 1404 cm⁻¹ and the $\nu_2(E)$ mode near 1665 cm⁻¹ (broken wavenumber axis), and (b) N–H stretching region, showing the main stretch near 3157 cm⁻¹ and the weaker band at 3238 cm⁻¹, stacked from 10 to 290 K. (c) Wavenumber (blue, left axis) and Lorentzian width $\Gamma_L$ (orange, right axis) of $\nu_4(F_2)$; (d) the same for $\nu_2(E)$; (e) the same for the main N–H stretch, with the solid red line showing the two-contribution fit of the position (Eq. 5) — the Einstein–Grüneisen $a$(T) contraction background (Eq. 4, green) plus the libron reversal. (f) Deviation δω(T) of the N–H-stretch position from the Einstein–Grüneisen contraction baseline; the solid line is the tanh libron fit (Eq. 3). The width in (c)–(e) is shown as data only: it is governed by dephasing, not by the anharmonic-plus-libron model (Sec. III C), and is not fitted.

The two ammonium bending modes shift only weakly with temperature (< 3 cm$^{-1}$ over the whole range): the $\nu_2$(E) mode near 1665 cm$^{-1}$ hardens on cooling, as expected when the N–H···Cl contacts strengthen and stiffen the H–N–H bend, whereas the $\nu_4$($F_2$) mode near 1404 cm$^{-1}$ shows a still weaker shift (≈ 1.3 cm$^{-1}$) of the opposite sign. The more informative temperature signal is in the line shape. The Lorentzian width of $\nu_2$(E) grows by a factor of ≈ 3.6 between 10 and 250 K, and the $\nu_4$($F_2$) profile evolves from Gaussian-dominated at low temperature to Lorentzian-dominated at high temperature — the classic signature of motional narrowing, in which the broadening changes from static-inhomogeneous at low temperature to dynamic-homogeneous as thermal modulation of the transition wavenumber becomes fast [42].

This broadening is not the ordinary anharmonic (Klemens) decay that describes the $[SnCl_6]^{2-}$ modes [43]. For a bending mode near 1400–1670 cm$^{-1}$ the down-conversion channel produces two phonons of ≈ 700–830 cm$^{-1}$, whose Bose occupation is negligible below 250 K; the corresponding two-phonon factor $[1 + 2n(\omega_0/2)]$ varies by less than 2 % over the measured range, so the anharmonic model predicts an essentially temperature-independent width. The strong, roughly linear growth of $\Gamma_L$ must therefore come from pure dephasing: elastic modulation of the high-wavenumber intramolecular vibration by a thermally populated low-energy mode, which randomizes the phase of the internal oscillation without relaxing its population [44,45]. In the exchange (or energy-modulation) picture, the dephasing rate scales with the occupation of the modulating mode, $\Gamma_{\text{deph}} \propto n(\Omega_{\text{low}})[\,n(\Omega_{\text{low}}) + 1\,]$; when $k_B T \gg \hbar\Omega_{\text{low}}$ this reduces to $n \approx k_B T/\hbar\Omega_{\text{low}}$, giving a width linear in temperature. The linearity we observe therefore points directly to a low-energy modulating mode — the librational and translational manifold of the $NH_4^+$ ion (Sec. III B, Sec. III C) — rather than to the internal mode itself. Because dephasing enters the width but leaves the line position almost unshifted, it accounts naturally for the pairing of a strongly temperature-dependent width with an almost temperature-independent centre, and it does not, on its own, define a libron energy; we therefore do not extract $E_{\text{eff}}$ from the bending-mode widths. Below ≈ 100 K the $\nu_2$(E) band develops a weak low-wavenumber satellite (near 1658 cm$^{-1}$), mirroring the crossover seen in the $[SnCl_6]^{2-}$ modes and in the N–H stretch, although its low intensity precludes a quantitative analysis.

The N–H stretching mode near 3157 cm$^{-1}$ is the most informative organic probe, because the X–H stretching wavenumber is a direct reporter of hydrogen-bond strength: a shorter, stronger N–H···Cl contact lengthens and weakens the covalent N–H bond and redshifts the stretch. On cooling from 300 K the mode softens to a minimum near 110–130 K and then hardens down to 10 K, tracing a shallow, asymmetric V — a ≈ 3.8 cm$^{-1}$ softening from 300 K to the minimum, followed by a ≈ 2.0 cm$^{-1}$ hardening below it (Figure 4b,e). This non-monotonic behaviour cannot be anharmonic in origin: for a 3157 cm$^{-1}$ mode the two-phonon Bose factor of the ≈ 1575 cm$^{-1}$ decay channel is ≈ $10^{-4}$ at 300 K, so the ordinary anharmonic wavenumber shift is frozen out over the whole range. The temperature dependence must therefore arise from the hydrogen bond and from the ammonium libration. A second, weaker N–H stretch is resolved at 3238 $cm^{-1}$ (Figure 4b); it loses intensity by roughly a factor of five on warming, but its temperature dependence is too weak to yield an independent libron energy and it is not analysed further.

The measured single-crystal $a$(T) is represented by a first-order Einstein–Grüneisen expression, which saturates as $T \to 0$ so that the thermal expansion coefficient vanishes at the absolute zero, as required by the third law of thermodynamics,

$$a(T) = a_0 + m_a \frac{\theta_E}{e^{\theta_E/T} - 1} \tag{4}$$

with $\theta_E \approx 180$ K (Figure 1b). The stretch position is then described by two contributions,

$$\omega(T) = \omega_0 + \kappa \, [\, a(T) - a_{\mathrm{ref}} \,] + S \tanh\left(\frac{E_{\mathrm{eff}}}{2k_B T}\right) \tag{5}$$

the first following the lattice contraction and the second being the same libron population factor used for the inorganic modes (Eq. 3). Here $a_{\mathrm{ref}} = a(10\ \mathrm{K})$ is a fixed reference value rather than a fitted quantity: it merely sets the temperature at which $\omega_0$ is defined and is entirely absorbed into $\omega_0$, so Eq. (5) has four free parameters, $\omega_0$, κ, $S$ and $E_{\mathrm{eff}}$. The contraction term reproduces the high-temperature softening — the H-bond strengthens as the cage contracts, with $\kappa \approx 52$ $cm^{-1}$ $Å^{-1}$; the same softening of the N–H stretch on cooling accompanies hydrogen-bond strengthening as the organic cation's rotation freezes in $MAPbI_3$ [46]. The libron term produces the low-temperature reversal through an angular mechanism. Hydrogen-bond strength depends on both the H···Cl distance and the N–H···Cl angle, and is greatest for a short, near-linear contact. As the cage contracts, the bonds first shorten while remaining close to linear (strengthening; redshift); once the ion is orientationally frozen below the crossover, further contraction is accommodated by a bending of the N–H···Cl angle away from linearity, which weakens the bond and reverses the shift (blueshift) even as the N···Cl distance keeps decreasing. The minimum near 110–130 K therefore marks the change of dominance between the two contributions, and it coincides with the $\nu(A_{1g})$ renormalization onset (≈ 90–100 K, Sec. III C) and with the classical-to-quantum crossover reported by INS and NQR [10,11].

Two features of this analysis deserve emphasis. First, the Lorentzian linewidth of the N–H stretch grows monotonically with temperature (Figure 4e), by a factor of ≈ 4 between 10 and 300 K, with no anomaly at the crossover. As for the ammonium bending modes (Sec. III C), this growth is not anharmonic — the Klemens two-phonon channel is frozen at 3157 $cm^{-1}$ — but pure dephasing by the low-energy librational and translational manifold; being a phase-relaxation ($T_2$) process it is monotonic and carries no crossover signature. The crossover therefore enters only the position, through the libron self-energy, and not the width. This is the opposite ordering to the low-frequency $[SnCl_6]^{2-}$ modes, whose two-phonon channel is thermally active: there the anharmonic width is itself temperature-dependent and the libron deviation is superimposed on it, so that both position and width track the crossover. Second, the effective energy $E_{eff}$ is not independently pinned by this single mode. With the background locked to the measured *a*(T) (only its amplitude κ free) and the full data set, the model fit returns $E_{eff} \approx 9.3$ meV, with a bootstrap FWHM range of 7.3–11.8 meV — a value that gives a markedly better fit than *either* the inorganic effective scale (≈ 4.7 meV) *or* the bare librational transition measured by INS (13.4 meV; [10]). As expected for a saturating term, the estimate is carried by the curvature of the low-temperature hardening below the ≈ 120 K minimum: restricting the fit to T ≳ 30 K, where that reversal is absent, lets $E_{eff}$ drift up toward the INS value. The robust statement is therefore that the N–H-stretch effective energy exceeds the inorganic effective scale, while its relation to the bare INS transition is not resolved. Read together, the two Raman probes and the neutron measurement suggest a hierarchy of energies — inorganic effective (≈ 4.7 meV, FWHM 3.5–6.2) < organic effective (≈ 9.3 meV, FWHM 7.3–11.8) < bare librational (13.4 meV) — in which the inorganic and organic bootstrap ranges do not overlap, so the ordering of the two Raman scales is supported by the resampling and not merely by the point estimates. This is consistent with the general expectation that a Raman self-energy scale lies below the bare transition, and with the N–H stretch, modulated directly by the libration of the same ion, coupling to it more strongly than the indirectly-coupled $[SnCl_6]^{2-}$ modes. This ordering is an interpretation of the effective energies, not a direct measurement of the librational spectrum.

The same hydrogen bond underlies the response to pressure (Sec. III D). Because both cooling and compression contract the cage, the N–H stretch shows an analogous V under pressure, with a minimum near 1.7 GPa (Sec. III D). The analogy is only partial, and the difference is informative: the thermal contraction between 250 and 130 K is ≈ 0.2 % in *a*, more than an order of magnitude smaller than the ≈ 2.7 % estimated for 1.7 GPa from the third-order Birch–Murnaghan equation of state of the isostructural $Rb_2TeBr_6$ ($K_0$ = 15.3 GPa, $B'$ = 7.7; [16]), no high-pressure diffraction having been measured here; lattice contraction alone cannot account for the thermal crossover — the librational freezing must be its dominant driver at ambient pressure. Pressure therefore acts predominantly on the static, hydrogen-bond geometry, whereas cooling convolves that geometry with the dynamical freezing of the rotor; comparing the two separates the contributions to first order (Sec. III D).

The absence of structural phase transitions allows the observation of a continuous crossover in the proton dynamics of the $[NH_4]^+$ ion — from thermally activated stochastic jumps over saddle points at high temperature to rotational tunneling at low temperature [10]. The low-temperature behavior of $(NH_4)_2SnCl_6$ was first explored by Prager *et al.* [10] by neutron diffraction. These authors observed a splitting of librational ground states, followed by classical reorientation with an activation energy of ≈590(30) K, in the same temperature range examined here, and described this as a continuous transition from classical diffusion to the quantum-mechanical limit for the $[NH_4]^+$ ion.

Subsequent temperature-dependent NQR and Raman studies attributed the broadening of internal $NH_4^+$ phonons to non-cubic interaction potentials between the $[SnCl_6]$ octahedra and the proton motions [12]. Fimland and Svare further reported dielectric losses associated with $[NH_4]^+$ motion, suggesting a local deviation from cubic symmetry — consistent with the mode asymmetries that emerge here below ≈100 K (Sec. III C), even though the average structure remains cubic with smoothly varying lattice parameters (Figure 1b,c). Neutron Bragg-scattering experiments by Brückel *et al.* [9] confirmed that, below 100 K, quantum effects are essential for a complete description of the low-temperature dynamics of $(NH_4)_2SnCl_6$.

The theory of Hüller and Kroll [13] models the librational states as linear combinations of products of quantum harmonic oscillators and nuclear spin states. The four-proton system has three distinct spin-function types (A, T, E), labeled by total nuclear spin $I$ = 2, 1, 0, respectively. As observed for the $[NH_4]^+$ ion in $(NH_4)_2PdCl_6$ [39] and for the methyl group in $(CH_3)_2SnCl_2$ [47], the coupling between lattice phonons and the totally symmetric librational ground state is markedly stronger than the coupling to the other two librational states. As temperature decreases, the population of the ground state increases, enhancing the effective coupling between librational modes and lattice vibrations — which we propose accounts for the anomalies observed in the Raman spectra of $(NH_4)_2SnCl_6$.

**D. Pressure-dependent Raman spectroscopy**

Hydrostatic compression has proved a powerful probe of low-dimensional halide perovskites, where it drives structural and optical transitions that are inaccessible at ambient pressure [48,49,50,51]. High-pressure Raman spectroscopy was therefore performed to probe lattice-phonon coupling arising from pressure-induced lattice distortion, up to a hydrostatic pressure of 10.1 GPa. Figure 5(a) shows the pressure-dependent spectra in the low-wavenumber region; all phonons in this region blueshift with increasing pressure, indicating hardening of the chemical bonds.

A subtle mode at 137 $cm^{-1}$ gains intensity from ≈1.3 GPa onward. This is the external translational $F_{2g}$ mode of $NH_4^+$ predicted by group theory (Sec. III B); it is the same weak translational mode already resolved near 120 $cm^{-1}$ at ambient pressure in the temperature series (Sec. III C), which is barely visible at room temperature and is progressively enhanced here by compression as it is by cooling. This low-wavenumber mode has previously been reported in other vacancy-ordered perovskites of the $R_2MX_6$ family [35]. While present under ambient conditions in compounds such as $Cs_2SnCl_6$ [34,52], isostructural compounds such as $Cs_2SnBr_6$ and $Cs_2SnI_6$ show this phonon only above 1 GPa [20]. The same behavior has recently been reported for the vacancy-ordered $Rb_2TeBr_6$, in which the A-site (Rb) translational $F_{2g}$ mode is undetectable at ambient pressure and becomes prominent above ≈1.6 GPa [16].

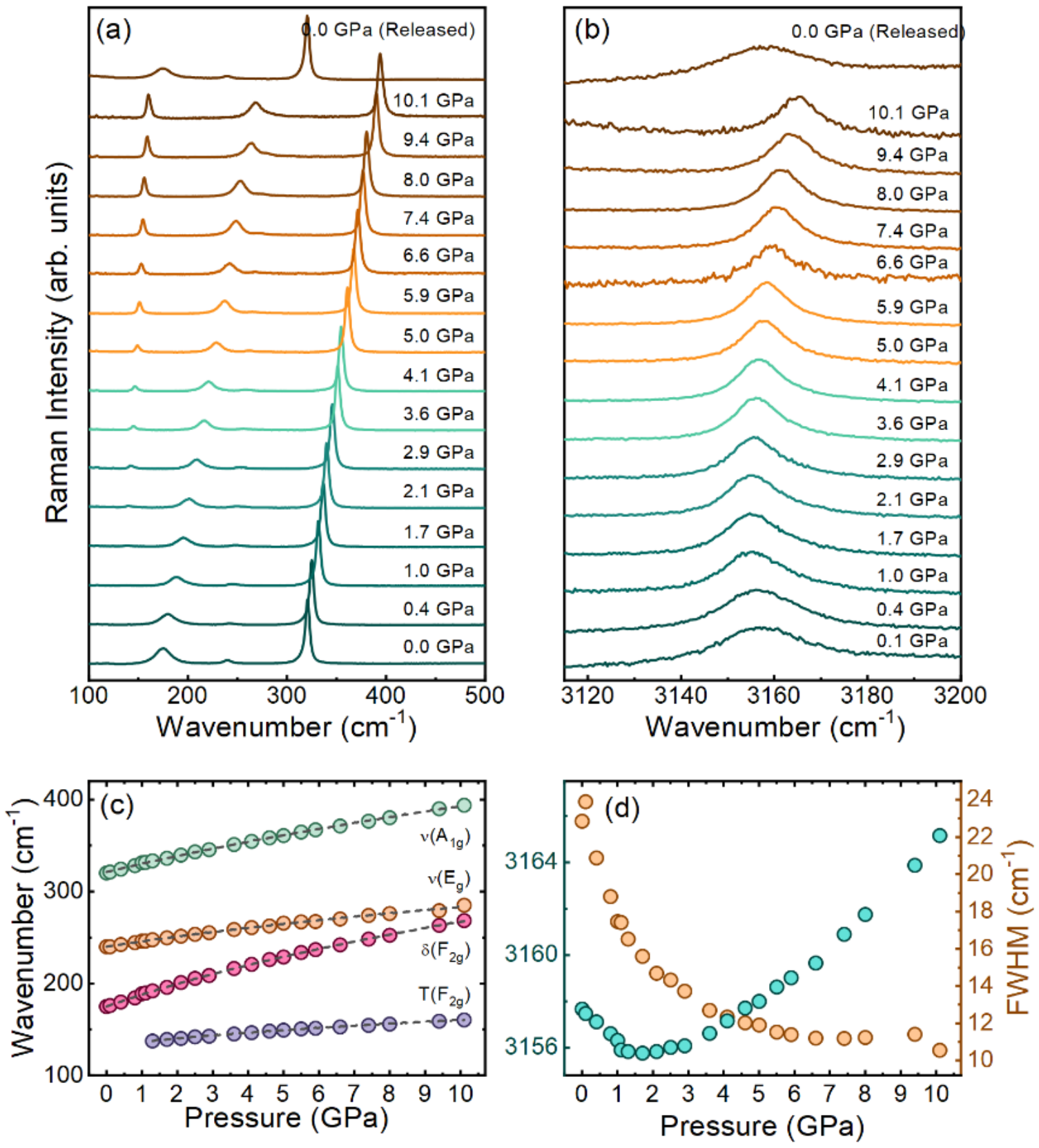

**Figure 5.** Pressure dependence of the room-temperature Raman spectrum of $(NH_4)_2SnCl_6$. (a) Low-wavenumber region (100–500 cm⁻¹) and (b) N–H stretching region (3120–3200 cm⁻¹), from ambient pressure to 10.1 GPa, with the spectrum recovered on decompression ("Released") shown at the top; the $\delta(F_{2g})$, $\nu(E_g)$ and $\nu(A_{1g})$ modes are labelled in (a). (c) Wavenumber versus pressure of the four low-wavenumber modes — $\nu(A_{1g})$, $\nu(E_g)$, $\delta(F_{2g})$ and the external translational $T(F_{2g})$ mode — with fits to Eq. (6) (lines). (d) Wavenumber (blue, left axis) and Lorentzian width (red, right axis) of the N–H stretch, showing the V-shaped minimum at which the pressure slope inverts.

All four low-wavenumber modes blueshift monotonically and sublinearly with pressure (Figure 5c). Sublinearity is expected rather than anomalous: in the quasi-harmonic approximation $\partial\omega/\partial P = \gamma\omega/K$, so that the slope must decrease as the bulk modulus K stiffens under compression. Combining this relation with a Murnaghan equation of state, $K(P) = K_0 + K'P$, gives

$$\omega(P) = \omega_0 \left(1 + \frac{K'P}{K_0}\right)^{\gamma/K'} \quad (6)$$

which reproduces every mode over the full range with residuals of 0.15–0.70 $cm^{-1}$ and returns the initial logarithmic pressure coefficient $\partial ln\omega/\partial P|_0 = \gamma/K_0$ directly. The fitted values are 79.8 ± 1.4, 33.6 ± 1.5, 29.0 ± 0.7 and 24.7 ± 1.3 × $10^{-3}$ $GPa^{-1}$ for $\delta(F_{2g})$, the external translational $F_{2g}$ mode, $\nu(A_{1g})$ and ν(Eg), respectively. None of the internal modes shows an anomaly at the pressure where the translational mode gains intensity; their sublinear stiffening is the ordinary consequence of the decreasing compressibility of the lattice, whose strain is taken up mainly by the interstitial cavity while the rigid octahedra are little affected [53]; in $Rb_2TeBr_6$ this partition has been resolved directly by high-pressure diffraction, the A-site cavity compressing markedly more than the $[BX_6]^{2-}$ octahedra, whose volume is almost unchanged over the first few GPa [16]. A comparable reduction of ∂ω/∂P within the cubic stability field, unaccompanied by any crystallographic transition, has been reported for $Rb_2TeBr_6$ and traced to the same mechanism, the decreasing compressibility of the octahedral subunits [16]. The $\delta(F_{2g})$ bend is thus by far the most pressure-sensitive mode, some three times more responsive than either stretch, consistent with a bending coordinate probing the softer Cl–Sn–Cl angles rather than the stiff Sn–Cl bonds. The external translational mode of $NH_4^+$ ranks second, as expected for a motion confined to the compressible cavity rather than to the rigid octahedron. Bulk moduli reported for the vacancy-ordered family span $K_0 \approx 11$ GPa for the ultrasoft $Cs_2SnBr_6$ [53] to 15.3 GPa for $Rb_2TeBr_6$ [16]; adopting the lower value as a conservative family estimate, these correspond to mode Grüneisen parameters $\gamma_i = K_0 \partial ln\omega/\partial P|_0$ of ≈ 0.9, 0.4, 0.3 and 0.3 for $\delta(F_{2g})$, $T(F_{2g})$, $\nu(A_{1g})$ and $\nu(E_g)$, respectively; the chloride is expected to be somewhat stiffer, so these are lower-bound estimates that a compound-specific equation of state for $(NH_4)_2SnCl_6$ would refine (cf. the $Cs_2SnX_6$ series, [20]).

In perovskites with connected octahedra, pressure-induced phase transitions generally arise from octahedral tilting; in 0D materials this mechanism is absent. Instead, pressure mainly compresses the interstitial space and shortens bond lengths rather than inducing tilt. The intensity gain of the external translational $F_{2g}$ mode at 1.3 GPa is thus the only feature localized at that pressure in the low-wavenumber spectrum. Because this mode is symmetry-allowed under the ambient $Fm\bar{3}m$ space group — corresponding to one of the four $F_{2g}$ modes predicted in Sec. III B — and merely gains Raman intensity as the cavity tightens around the ammonium ion, rather than being created by a lowering of symmetry, its appearance does not signal a structural phase transition. Consistently, none of the internal modes splits or shows an anomaly over the whole pressure range, and the ambient spectrum is fully recovered on decompression (Figure 5a–b). We therefore describe the 1.3 GPa feature as a continuous, isosymmetric onset within $Fm\bar{3}m$, not a phase transition [20]. This is consistent with the known thermal behaviour of the compound: adiabatic calorimetry of $(NH_4)_2SnCl_6$ shows only the smooth, anomaly-free heat capacity of hindered ammonium rotation, with no λ-anomaly [54], and its $^{35}Cl$ NQR frequency tracks the ammonium reorientation continuously from 4 to 450 K without a discontinuity [12] — unlike the bromide analogue and several alkylammonium hexachlorostannates, which do undergo temperature-driven transitions. A metastable orthorhombic polymorph of $(NH_4)_2SnCl_6$ has been grown from the vapour phase [55]; our crystals, however, remain cubic throughout (Sec. III A) and recover fully on decompression, so this polymorph is not accessed under the present temperature or pressure conditions.

This partition — rigid octahedra stiffening smoothly, while the cavity subsystem responds through both the ammonium translational mode (gaining intensity from ≈ 1.3 GPa) and the inversion of the N–H-stretch slope (near 1.7 GPa) — mirrors the temperature behavior, where the crossover is likewise carried by the ammonium/cage subsystem rather than the octahedra, and underscores the 0D, molecular-crystal character of the compression.

Because compression in such materials occurs primarily within the interstitial space, the ammonium internal modes provide a valuable probe of the effect of hydrostatic pressure on $(NH_4)_2SnCl_6$. The pressure-dependent ammonium spectra are shown in Figure 5(b), with the pressure dependence of the symmetric-stretching parameters in Figure 5(d).

The ammonium bending region, however, could not be followed under pressure, for two distinct reasons. The $\nu_4(F_2)$ mode (≈ 1404 $cm^{-1}$) lies on the shoulder of the intense first-order diamond Raman line (≈ 1332 $cm^{-1}$) of the diamond-anvil cell, which dominates that spectral window; the $\nu_2(E)$ mode (≈ 1665 $cm^{-1}$), although well separated from the diamond line, is intrinsically weak (Sec. III C); it is occasionally discernible within the cell but is too weak to be reliably fitted under its reduced signal-to-noise. The pressure analysis of the ammonium subsystem is therefore restricted to the N–H stretching region.

As for the inorganic modes, no discontinuity is detected in phonon position or width; the sign of $\partial\omega/\partial P$ of the N–H stretch, however, inverts near 1.7 GPa, close to but not coincident with the pressure at which the $F_{2g}$ translational mode gains intensity (1.3 GPa). Below 1.7 GPa the stretch softens as compression shortens and strengthens the N–H···Cl bonds ($\Delta\omega \approx -2.0$ $cm^{-1}$ from ambient); above it the mode hardens as the bonds, having reached their shortest length and widest angle, begin to bend rather than shorten further [56] — the pressure counterpart of the low-temperature reversal (Sec. III C). The two routes are not identical, however. Compression, like cooling, tightens the cavity and is expected to raise the barrier to ammonium reorientation — at room temperature the $NH_4^+$ ion reorients almost freely [11,12], and pressure should drive it toward the hindered, librational regime. The linewidth supports this: the N–H-stretch FWHM collapses from ≈ 22 to ≈ 7 $cm^{-1}$ between ambient and 10 GPa, and since that width is set by dephasing against the low-energy librational and translational manifold (Sec. III C), its narrowing signals a loss of orientational disorder as the rotor localizes. Compression therefore acts on the cavity through two channels — the static hydrogen-bond geometry, which dominates the N–H-stretch reversal, and a partial hindering of the rotor dynamics — that we cannot fully separate without pressure-dependent diffraction or quasi-elastic neutron scattering, neither of which is available here. That the octahedral modes stiffen smoothly while the two cage-related features — the translational-mode intensity onset and the N–H-stretch reversal — sit close together but at slightly different pressures confirms that the high-pressure response is confined to the cage, without implying a single shared threshold. Comparing the pressure and temperature data therefore separates, to first order, the static hydrogen-bond-geometry contribution — dominant under pressure — from the dynamical, librational one that dominates on cooling; the separation is not exact, since compression also stiffens the rotor, but it remains the central advantage of the combined temperature/pressure dataset.

## IV. CONCLUSIONS

The vacancy-ordered perovskite $(NH_4)_2SnCl_6$ was synthesized and its structure and lattice dynamics followed as a function of temperature and pressure by single-crystal X-ray diffraction and Raman spectroscopy. No structural phase transition occurs between 300 and 10 K, and the average cubic lattice parameters vary smoothly with no anomaly (Figure 1b,c). Below ≈100 K the Raman spectra nonetheless develop local, dynamical signatures: the $[SnCl_6]^{2-}$ modes acquire a libron–phonon renormalization and a growing, symmetry-selective asymmetry, and the N–H stretching mode passes through a shallow minimum near 110–130 K. Both track the classical-to-quantum crossover of the ammonium rotor established by neutron scattering and NQR, now observed through its coupling to the Raman-active phonons rather than through any change in the average structure — a dynamical rather than structural crossover.

The N–H stretch serves as a hydrogen-bond gauge: it softens as the cage contracts on cooling, when the N–H···Cl bonds shorten and strengthen, and reverses once the rotor localizes and the bonds bend rather than shorten further. Its temperature dependence cannot be anharmonic — the two-phonon decay channel is frozen out at this wavenumber — so it reports the hydrogen bond and the libration directly. Its linewidth, by contrast, is governed by pure dephasing against the low-energy librational and translational manifold, as are the ammonium bending modes, and carries no crossover signature. The effective librational energy the position returns (≈ 9.3 meV) exceeds the ≈ 4.7 meV scale that renormalizes the $[SnCl_6]^{2-}$ modes, consistent with the N–H stretch coupling more directly to the same libration, although the single mode does not fix it precisely and its relation to the 13.4 meV neutron transition is not resolved.

Under pressure the three internal $[SnCl_6]^{2-}$ modes stiffen monotonically and sublinearly, with no anomaly at 1.3 GPa: their decreasing pressure coefficients reflect the ordinary stiffening of the lattice, not a transition. The cage-related features belong instead to the ammonium/cavity subsystem — the symmetry-allowed external translational $F_{2g}$ mode gains Raman intensity from ≈ 1.3 GPa as the cavity tightens, and the N–H stretch inverts its pressure slope near 1.7 GPa as the hydrogen bonds pass from shortening to bending. Neither breaks the cubic symmetry: the internal modes neither split nor show an anomaly, and the spectrum recovers fully on decompression. The 1.3 GPa feature is therefore a continuous, isosymmetric onset within $Fm\bar{3}m$, not a phase transition.

Temperature and pressure therefore provide complementary views of the same underlying mechanism. The rigid $[SnCl_6]^{2-}$ octahedra are largely inert — they stiffen smoothly under compression and renormalize only weakly on cooling — while the ammonium subsystem, through its libration and its hydrogen bonds, carries the response to both perturbations. Because compression acts predominantly on the hydrogen-bond geometry while cooling both contracts the cage and freezes the rotor, the two experiments together separate, to first order, the static contribution from the dynamical one. This partition of the lattice dynamics into a rigid inorganic framework and a dynamically active molecular sublattice is characteristic of the 0D vacancy-ordered structure, and is the dynamical counterpart of the two-tier vibrational bath recently proposed for $(NH_4)_2SnCl_6$, in which $[SnCl_6]^{2-}$ phonons drive the self-trapping of photoexcited carriers while $NH_4^+$ rotational and librational motions modulate the vibronic relaxation and broaden the emission [23]. The measurements reported here place an energy scale on the coupling between the two subsystems and suggest a route to tune cation–phonon coupling independently of the octahedral network.

## ACKNOWLEDGMENTS

This study was partially financed by the Brazilian agencies CAPES (Finance Code 001), FUNCAP (PRONEX PR2-0101-00006.01.00/15, Rede Verdes 07548003/2023), and CNPq (Grants No. 407954/2022-8, 407956/2022-0, INCT MatFerrCE 406322/2022-8).

## DATA AVAILABILITY

The data that support the findings of this study are available from the corresponding author upon reasonable request.

# Libron–phonon coupling and hydrogen-bond dynamics in the vacancy-ordered perovskite $(NH_4)_2SnCl_6$: a temperature- and pressure-dependent Raman study

Vasco S. Neto[1], Mayra A. P. Gómez[1,2], Bruno S. Araújo[1], Alejandro P. Ayala[1]*

[1] Departamento de Física, Universidade Federal do Ceará, Campus do Pici, Fortaleza, CE 60440-900, Brazil

[2] Departamento de Física dos Materiais e Mecânica (FMT), Instituto de Física da Universidade de São Paulo (IF-USP), São Paulo, São Paulo, Brasil. CEP 05314-970.

## SUPPLEMENTAL MATERIAL

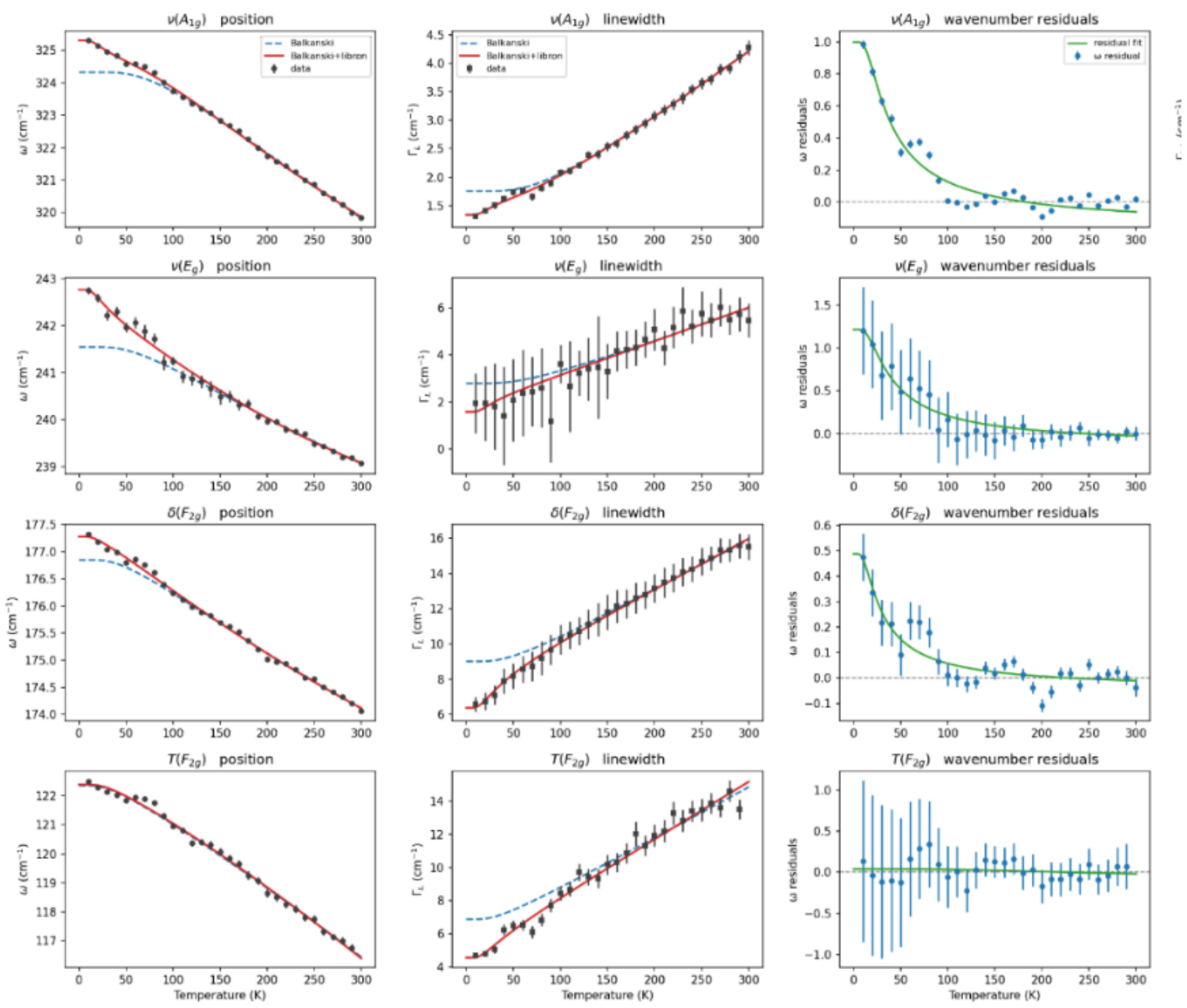

**Figure S1.** Complete anharmonic-plus-libron analysis of the four low-wavenumber Raman modes: $\nu(A_{1g})$, $\nu(E_g)$, $\delta(F_{2g})$, and the external translational $F_{2g}$ mode of $NH_4^+$ (labelled $T(F_{2g})$ in Figure 2 of the main text). Columns: wavenumber $\omega(T)$ (left), Lorentzian width $\Gamma_L(T)$ (middle), and residual $\omega(T) - \omega_{anharmonic}(T)$ (right). Dashed blue line, anharmonic (Balkanski) model; solid red line, anharmonic-plus-libron model; solid green line, tanh fit to the residual (Eq. 3). All four modes show the same qualitative deviation from the purely anharmonic baseline below ≈100 K, though the effect is weakest and least resolved for the translational $F_{2g}$ mode, consistent with its assignment as a rigid-body cation translation that couples only weakly to the $NH_4^+$ libration (Sec. III C).

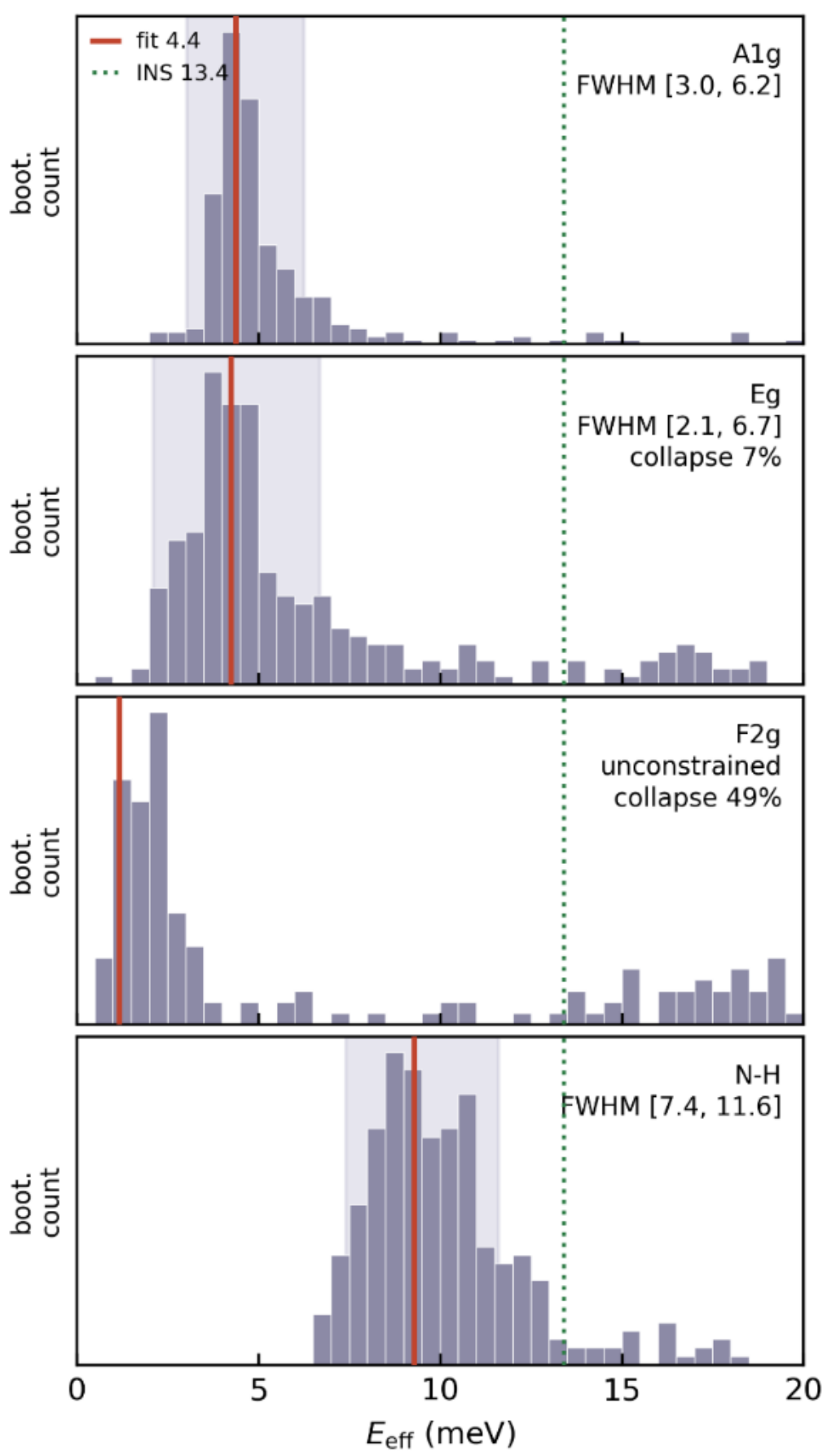

**Figure S2.** Bootstrap distributions of the effective librational energy $E_{eff}$ for the internal $[SnCl^6]^{2-}$ modes ($\nu(A_{1g})$, $\nu(E_g)$, $\delta(F_{2g})$) and the N–H stretch, obtained by case resampling (1000 replicates); the complete fitting procedure — alternating backfit, multistart over several seeds for $E$, and a final joint refinement — was re-run independently on each replicate, so that the spread reflects sampling variability rather than the local curvature of $\chi^2$. Red line, $E_{eff}$ from the fit to the full data set, also quoted in each panel; green dotted line, the bare 0→1 librational transition from INS (13.4 meV [10]). Shaded band, full width at half maximum of a Gaussian kernel-density estimate of the distribution, quoted as the uncertainty range; it measures the width of the distribution and is not a confidence interval. For $\nu(A_{1g})$, $\nu(E_g)$ and the N–H stretch the distributions are single-peaked and the point estimate falls within the band. For $\delta(F_{2g})$ the libron amplitude is too small for $E_{eff}$ to be determined: the fit collapses in ≈32 % of replicates, the surviving distribution is bimodal, and no band is drawn. All distributions are skewed towards high $E_{eff}$ because the energy is fixed by the curvature of the low-temperature saturation of Eq. (3), so replicates that sample that region sparsely return large values. This asymmetry is intrinsic to a saturating two-level term rather than a limitation of the measurement, and is why the FWHM rather than a percentile interval is quoted in the text.